\documentclass{icrc2009}

\usepackage{graphicx}   % for including figures
\usepackage[caption=false]{caption}    % for captions
\usepackage[font=footnotesize]{subfig} % subfig.sty for a double column floating figure using two subfigures
\usepackage{fixltx2e}
\usepackage{url}

\newcommand{\shorttitle}[1]%
{\markboth{Proceedings of the 31\MakeLowercase{$^{st}$} ICRC, {\L}\'{o}d\'{z} 2009}{#1} }
\newcommand{\etal}{\MakeLowercase{\textit{et al. }}} % "et al."

\usepackage{amsmath}
\begin{document}
\title{Search for Cosmic-Ray Antiparticles with Balloon-borne Experiments}

\author{\IEEEauthorblockN{Ph. von Doetinchem\IEEEauthorrefmark{1},
			  H. Gast\IEEEauthorrefmark{1} and
                          St. Schael\IEEEauthorrefmark{1}}
			  \\
\IEEEauthorblockA{\IEEEauthorrefmark{1}I. Physics Institute B, RWTH Aachen University, Sommerfeldstr. 14, 52074 Aachen, Germany}
}

% please write the preseter's name and short title (3-4 words maximum)
%    which will appear at the header of the even pages.
\shorttitle{Ph. von Doetinchem \etal Antiparticles with Balloon-borne Experiments}
\maketitle

\begin{abstract}
This work discusses the prospects of antiparticle flux measurements with the proposed PEBS detector. The project foresees long duration balloon flights at one of Earth's poles at an altitude of 40\,km. The sky coverage of flights at the poles is presented. In addition, cosmic-ray measurements at the poles (small rigidity cut-offs) give the possibility to study solar modulation effects down to energies of about 0.1\,GeV. Furthermore, systematic effects due to interactions of cosmic rays in the atmosphere are important. These effects were studied with the Planetocosmics simulation software based on GEANT4 in the energy range 0.1 - 1000\,GeV.
  \end{abstract}


\section{The PEBS Experiment}

The Positron Electron Balloon Spectrometer (PEBS) is a proposal for a balloon-borne experiment to fly at an altitude of about 40\,km in the atmosphere of the Earth \cite{doe07,gas09}. The long duration flights are planned for the North and the South Pole. The poles have several advantages: the flight latitude stays quite stable and the landing position is predictable due to the stable circumpolar winds. The measurements must take place in summer to have good accessibility of the launching and landing sites and to be able to generate the power with solar panels. The total measuring time of several flights is planned to add up to 100\,days.

The PEBS detector will consist of several subsystems with a total acceptance of 0.4\,m$^2$sr. A time of flight system (TOF) with one lower and one upper plane is needed for triggering and for velocity measurements. Two transition radiation detectors (TRD) are used for the discrimination between light and heavy charged particles. A combined silicon and scintillating fiber tracker is used for momentum measurement (resolution: $\sigma_p/p=0.02\,\%\cdot p/\text{GeV}\oplus2.3\,\%$) and a sandwich electromagnetic calorimeter (ECAL) with embedded fibers for the discrimination of particles. The detector is located inside a superconducting magnet with an average field of 0.8\,T. 

\begin{figure}
\centering
\includegraphics[width=1.0\linewidth]{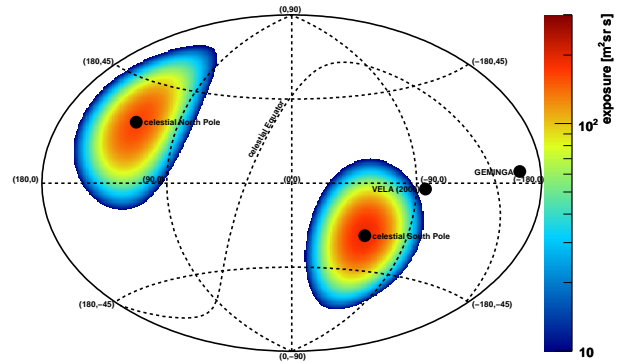}
\caption{Exposure of the sky in galactic coordinates (Aitoff projection) for the balloon-borne PEBS experiment with an acceptance of 0.4\,m$^2$sr (maximum opening angle $\approx60^\circ$) at the North Pole and South Pole for a flight time of 50\,days each.}
\label{f-northsouthpole_longitude_latitude}
\end{figure}

\begin{figure}
\centering
\includegraphics[width=1.0\linewidth]{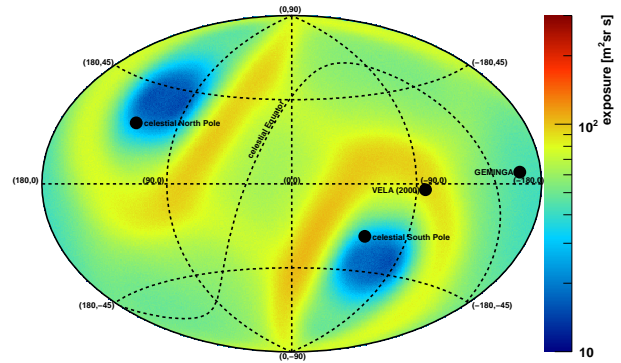}
\caption{Exposure of the sky in galactic coordinates (Aitoff projection) for the space-based AMS-02 experiment with an acceptance of 0.095\,m$^2$sr (maximum opening angle $\approx50^\circ$) at ISS orbit for a flight time of 3\,years.}
\label{f-issorbit_longitude_latitude}
\end{figure}
 
\section{Sky Coverage}

Recent data of the ATIC \cite{atic}, HESS \cite{hess} and FERMI \cite{fermi} experiments indicate an excess of the combined electron and positron flux above the background expectation. Several interpretations of these data are discussed in the literature, e.g. dark matter annihilations and pulsars \cite{hall-2008}. Therefore, it is interesting to map the galactic sky with charged particles in the TeV range even if they are deflected in the galactic magnetic field. It might still be possible to find anisotropies due to local sources. This could be used to distinguish between dark matter and pulsar models. We have simulated an isotropic particle distribution on the galactic sky and determined the exposure for each detector position. Here, magnetic fields are neglected and particles are assumed to follow straight lines. In comparison to PEBS, the sky coverage for the AMS-02 experiment \cite{ams04} mounted on the International Space Station is calculated.

The exposures are shown in fig.~\ref{f-northsouthpole_longitude_latitude} for PEBS with two flights (North and South Pole, 50\,days each) and in fig.~\ref{f-issorbit_longitude_latitude} for AMS-02 at ISS orbit for three years. The exposure is defined as the total number of entries normalized to the detector acceptance multiplied by the measurement time. The measurements with PEBS cover the regions around the celestial poles while AMS-02 covers nearly the complementary region. Only both experiments together are able to deliver a complete picture of the sky, e.g. of the nearby Vela Pulsar. These pictures can be interpreted as the expected diffuse charged particle background. A deeper study of possible new sources would have to take into account effects of the galactic magnetic field, e.g. as a function of distance to the source.

\section{Geomagnetic Cut-Offs}

\begin{figure}
\centering
\includegraphics[width=1.0\linewidth]{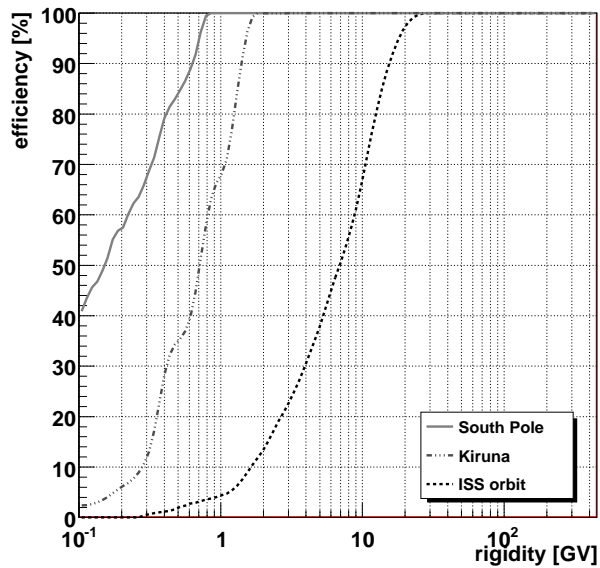}
\caption{Comparison between the efficiencies for geomagnetic cut-offs at South Pole, Kiruna and at ISS orbit.}
\label{f-cutoff_issorbit}
\end{figure}

The discrepancy between the PAMELA and the HEAT/AMS positron fraction data in the energy range below 4\,GeV \cite{pamela_ep} might result from different solar magnetic polarities which may influence the behavior of solar modulation \cite{gas09,gas_icrc}. The influence of the geomagnetic field should be small to study this effect. In general, deflection of particles is strong at the equator and weak at the poles depending additionally on rigidity. Therefore, the probability for a certain cut-off as a function of detector trajectory at the poles and at ISS orbit were calculated by tracing particles back to the outside of Earth's magnetosphere using PLANETOCOSMICS \cite{laurent-2005} based on GEANT4 \cite{geant4-1,geant4-2}. The cut-off efficiencies averaged over all detector positions and isotropic direction angles at the South Pole, Kiruna (close to the North Pole) and ISS orbit are shown in fig.~\ref{f-cutoff_issorbit}. The dependence on the geographic position is obvious and shows that PEBS will be able to measure cosmic rays to lower energies than AMS-02 which is important to investigate solar modulation.

\section{Influence of the Atmosphere\label{s-atmo}}

It is important for balloon experiments to understand the interactions of cosmic-ray particles with Earth's atmosphere. At 40\,km the atmospheric depth at the poles during summer is about 3.8\,g/cm$^2$. This is a non-negligible effect compared to the mean amount of matter traversed in the galaxy before entering the atmosphere (6 - 10\,g/cm$^2$). Particles with 0$^\circ$ zenith angle traverse about 10\,\% of a radiation length before reaching 40\,km. Thus, the particles resulting from electromagnetic atmospheric interactions can be interpreted to be most likely produced in the first interaction of the cosmic-ray particle with the atmosphere. This holds also for hadronic cascades as the corresponding nuclear mean free pathlength is about half of the radiation length in air. Particles from secondary interactions contribute to the measured fluxes while the primary cosmic-ray particles are attenuated. This can falsify the interpretation of the data. The software package PLANETOCOSMICS is used in the following to study these effects in detail.

\subsection{Cosmic-Ray Simulation in the Atmosphere}

The analysis starts with the calculation of cosmic-ray spectra with the GALPROP reacceleration model \cite{strong-1998-509,strong-2001-27,2006ApJ...642..902P}. In addition, the cosmic-ray fluxes must be solar modulated. Starting positions and directions for particle trajectories at an altitude of 500\,km above Earth's surface are calculated without the atmospheric and geomagnetic effects such that an isotropic particle flux would be achieved in detection shells around the Earth. Then the atmosphere and magnetic field is switched on for the simulation to study their effects. The simulation uses current models for the atmospheric composition NRLMSISE00 \cite{picone-2002} and the magnetic field IGRF \cite{igrf}. Protons, electrons, positrons, photons and antiprotons are simulated with energies between 0.1 and 450\,GeV for New Mexico and between 0.1 and 10,000\,GeV at the South Pole. The simulation of helium nuclei works only stable up to about 7.5\,GeV per nucleon due to the physics models implemented in GEANT4 for light ions.

\subsection{Comparison between Simulations and Measurements}

\begin{figure}
\centering
\includegraphics[width=1.0\linewidth]{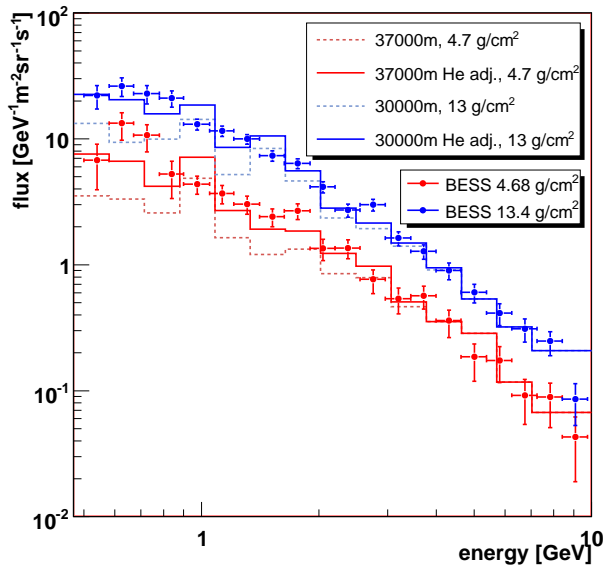}
\caption{Comparison of simulated and measured muon fluxes.}
\label{f-fluxes_compare_13_ftsumner}
\end{figure}

A comparison of the atmospheric muon flux measured by the BESS experiment \cite{abe-2007-645} for two different atmospheric depths with the fluxes simulated for New Mexico is shown in fig.~\ref{f-fluxes_compare_13_ftsumner}. The dashed lines show the muon flux from the simulation without adjustment. For both depths the simulation lies below the data at lower energies. This is because GEANT4 works stably only up to a few GeV per nucleon for helium nuclei. Therefore, the secondary muon flux caused by helium is too small. A correction for this effect is estimated by comparing cosmic proton induced muon fluxes of different energy ranges. The proton flux in the energy range of 0.1 - 30\,GeV (30\,GeV protons $\approx$ helium nuclei with 7.5\,GeV per nucl.) contributes about 20\,\% of the total atmospheric muon flux. Therefore, all secondary particle fluxes induced by helium are multiplied by 5 in the following. This effect is important for particle energies up to about 5\,GeV. The adjusted fluxes (solid lines) at 30\,km (13.0\,g/cm$^2$) and 37\,km (4.7\,g/cm$^2$) agree on average within about 15\,\% with the measurement. The traversed atmospheric depth increases with larger zenith angles and a measurement of the zenith angle distribution of muons would be an interesting test of atmospheric models.

\begin{figure}
\centering
\includegraphics[width=1.0\linewidth]{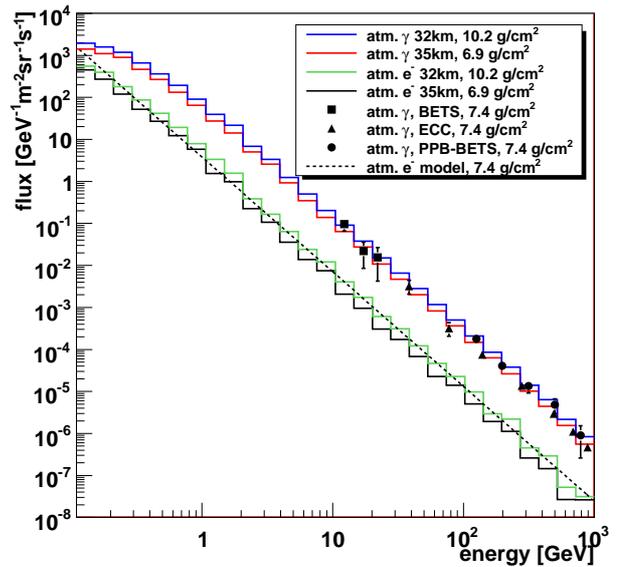}
\caption{Comparison of photon flux measurements and simulations and comparison of secondary electron flux simulations and an analytical model \cite{nishimura-1980,yoshida-2006,torii-2008}.}
\label{f-electron_gamma_compare}
\end{figure}

In addition, the comparison of the observed photon spectra at an atmospheric depth of 7.4\,g/cm$^2$ at the South Pole with the simulations at 32\,km (10.2\,g/cm$^2$) and 35\,km (6.9\,g/cm$^2$) at the South Pole shows good agreement within the error bars (fig.~\ref{f-electron_gamma_compare}). The figure shows also a model for the atmospherically induced electron flux used for the PPB-BETS experiment in comparison to the simulated electron fluxes from atmospheric interactions \cite{nishimura-1980,yoshida-2006,torii-2008}. The agreement between the model and the simulation is good. These comparisons are used for an estimation of the systematic uncertainty of the atmospheric simulations for the following analysis. The systematic error is assumed to be 15\,\%.

\subsection{Particle Fluxes at the South Pole}

Following the good agreement with the atmospheric data, the simulation at the South Pole for the PEBS experiment were carried out. The atmospheric depth is nearly the same in the arctic or antarctic summer at the corresponding pole and the geomagnetic cut-offs are very small at both poles. Therefore, further simulations have been carried out for the South Pole.

\begin{figure}
\centering
\includegraphics[width=1.0\linewidth]{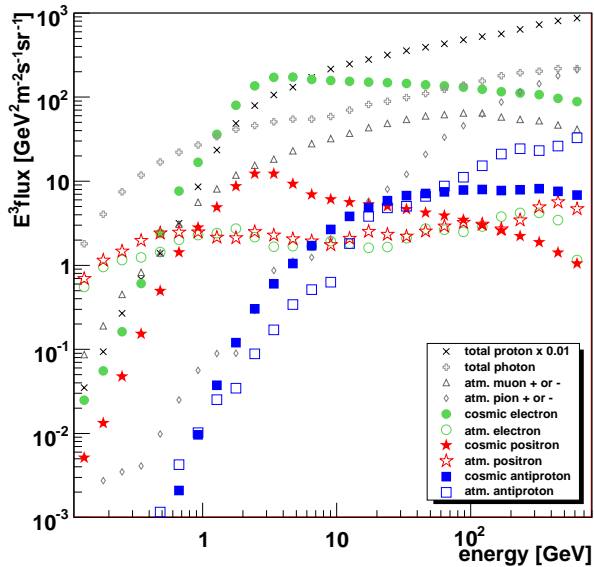}
\caption{Fluxes (smoothed) at 40\,km altitude (3.8\,g/cm$^2$) at the South Pole respecting the detector acceptance. No additional signals due to dark matter or pulsars are shown.}
\label{f-total_flux_40km}
\end{figure}

Fig.~\ref{f-total_flux_40km} shows for electrons, positrons and antiprotons the cosmic and atmospheric energy differential fluxes separately and for the other particle types the sum of cosmic and atmospheric contributions at 40\,km respecting PEBS angular detector acceptance such that most particles come directly from above. Cosmic antiprotons and positrons show non-negligible atmospheric backgrounds. The separation of cosmic photons from the huge amount of atmospheric photons is not possible. Therefore, photon measurements are not further discussed. Muons and pions arise only from interactions in the atmosphere and their large abundance require a good discrimination against them, e.g. using an ECAL or ring image \v{C}erenkov detector. Especially the antiproton measurement at higher energies will be limited due to large backgrounds of muons and pions. The challenge is obviously to correct for atmospheric effects to extract the cosmic-ray fluxes.

An important difference between antiprotons and positrons exists in their attenuation in the atmosphere. The energy loss of positrons in the atmosphere is stronger than for antiprotons. This is because positrons lose energy strongly due to bremsstrahlung and are shifted to lower energies. In comparison, the energy loss of antiprotons mostly due to nuclear interactions is smaller in the atmosphere. The atmospheric background flux of antiprotons below 1\,GeV exceeds slightly the cosmic flux. At about 60\,GeV the contribution to the total antiproton flux by atmospheric antiprotons is equal to the cosmic contribution and at 1\,TeV the atmospheric flux is about 5 times as large as the cosmic flux. The secondary production of positrons below 1\,GeV is much stronger than for antiprotons and exceeds the cosmic positrons on top of the atmosphere below 1\,GeV and makes a reliable measurement of the cosmic positron flux very difficult. The atmospheric and cosmic contribution are equal at about 0.5\,GeV for electrons and positrons. The simulations predict a change of the slope at this point and it would be interesting to measure this shoulder precisely to constrain the atmospheric model. The atmospheric positron production up to energies of 100\,GeV is dominated by muon decay and the atmospheric to cosmic flux ratio at 40\,km altitude is about 10 - 20\,\%. Neutral pion decay to photons followed by electron-positron pair production becomes important from about 100\,GeV where the cosmic and atmospheric positron flux is nearly equal.

\subsection{Flux Measurements with PEBS}

The total measured number of particles of a certain type at a certain altitude is influenced by interactions in the atmosphere and by detector misidentification. Misidentification becomes especially important if the background exceeds the signal as in the case of cosmic-ray particles and antiparticles. The cosmic-ray flux extraction takes into account an energy dependent rejection of protons and momentum resolution. Systematic uncertainties of 10\,\% for the detector effects are assumed while the systematic error for the atmospheric effects is set to 15\,\%. A deeper discussion of the detector and atmospheric properties can be found in \cite{gas09,doe09}.

\begin{figure}
\centering
\includegraphics[width=1\linewidth]{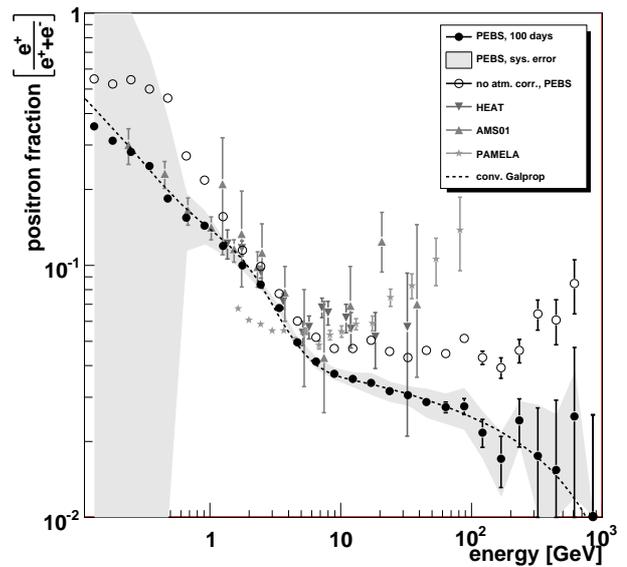}
\caption{Projected positron fraction, background expectation only \cite{pamela_ep,Barwick-1997,ams01,olzem-2007}.}
\label{f-e+_e-_fraction_dm_hg_phi550_40000_m_southpole_100_days}
\end{figure}

The resulting corrected cosmic positron fraction is shown in fig.~\ref{f-e+_e-_fraction_dm_hg_phi550_40000_m_southpole_100_days} together with the systematic error band composed of detector and atmospheric effects at 40\,km altitude. The positron fraction will be measured precisely starting from 1\,GeV up to about 200 - 300\,GeV. The positron fraction shows at lower energies large systematic errors due to large atmospheric corrections resulting mainly from strong atmospheric positron production and from bremsstrahlung losses of the cosmic positrons in the atmosphere. The open circles in fig.~\ref{f-e+_e-_fraction_dm_hg_phi550_40000_m_southpole_100_days} show the positron fraction without atmospheric corrections. It is clearly seen that the fraction would be overestimated without corrections especially for low and high energies.

The statistical limitations at higher energies might play a minor role if the PAMELA data are correct and the fraction shows an increase. This emphasizes the need of a well calibrated detector up to high energies for a reliable measurement.

\section{Conclusion}

It was shown that the proposed PEBS detector would give a higher accuracy than all experiments completed so far for cosmic-ray antiparticle measurements. These data would allow to test solar modulation models to better constrain the electron and positron fluxes at low energies. At higher energies PEBS could be used to test isotropy of the electron and positron fluxes to distinguish between the various proposed sources for the origin of the new features.

\end{document}